\documentclass[12pt]{article}

\usepackage{a4wide}
\usepackage{epsfig}
\usepackage{graphicx}
\usepackage{amsmath}
\usepackage{amsfonts,amssymb}
\usepackage{color}
\usepackage{float}

\newcommand{\pfrac}[2]{\left(\frac{#1}{#2}\right)}
\newcommand{\Tr}{{\rm Tr}}
\newcommand{\sn}{\mathop{\rm sn}\nolimits}
\newcommand{\MeV}{\mathop{\rm MeV}\nolimits}
\newcommand{\GeV}{\mathop{\rm GeV}\nolimits}
\newcommand{\bbbone}{\hbox{\rm 1\kern-3pt l}}
\newcommand{\real}{\mathop{\rm Re}\nolimits}
\newcommand{\dAlem}{\hbox{\,\vbox{\hrule height0.3pt\hbox{\vrule width0.3pt
  \hbox{\vbox to8pt{\hbox to8pt{\hfill}\vfill}}\vrule width1.0pt}\hrule
  height1.0pt}\,}}

\begin{document}

\begin{center}
{\Large\bf 
Finite temperature QCD crossover at non-zero chemical potential:
{\it A Dyson--Schwinger approach}
}
\\[12pt]
{\large
Arpan Chatterjee$^1$, Marco Frasca$^2$,\\
Anish Ghoshal$^3$ and Stefan Groote$^1$}\\[12pt]
$^1$ F\"u\"usika Instituut, Tartu Ulikool,
W.~Ostwaldi 1, EE-50411 Tartu, Estonia\\
$^2$ Independent Researcher, 00176 Rome, Italy\\
$^3$ Faculty of Physics, University of Warsaw,
ul.~Pasteura~5, 02-093 Warsaw, Poland





\end{center}
\vspace{1cm}
\begin{abstract}
\noindent
We study QCD at finite temperature and non-zero chemical potential to derive
the critical temperature at the chiral phase transition (crossover). We solve
a set of Dyson--Schwinger partial differential equations using the exact
solution for the Yang--Mills quantum field theory based on elliptical
functions. We derive a Nambu-Jona--Lasino (NJL) model of the quarks and obtain
a very good agreement with recent lattice computations regarding the
dependence of the critical temperature on the strong coupling scale. The
solution depends on a single scale parameter, as typical for the
theory and already known from studies about asymptotic freedom. The
study is analytically derived from QCD. 
\end{abstract}


\newpage

\section{Introduction}
Undergoing collisions of large nuclei at the Relativistic Heavy Ion Collider
(RHIC) at the Brookhaven National Laboratory (BNL) and the Large Hadron
Collider (LHC) create a hot plasma of quarks and gluons with the
properties of the early Universe, attaining a very large temperature and
involving almost approximate symmetry between matter and antimatter.
Future prospects at the FAIR experiment will be able to probe QCD phase
diagram structure and to understand the chiral and deconfinement transitions
from the hadronic state of matter to the quark-gluon plasma
phase~\cite{Chen:2015ona}. The upcoming devices like the sPHENIX
detector~\cite{Okawa:2023asr}, along with forseen complementary
STAR upgrades at RHIC~\cite{Tarnowsky:2011zza}, and together with the plans
for the higher luminosity run at the LHC augmented with already present and
future upgraded detectors at
ALICE~\cite{Cifarelli:2024kgh}, ATLAS~\cite{ATLAS:2023fjw},
CMS~\cite{Velkovska:2012fh} and LHCb~\cite{Vagnoni:2025vmf}, will give us the
opportunity to investigate such thermodynamics of QCD with the joint analysis
of data from low-energy hadrons, jets, thermal electromagnetic radiation of
plasma, heavy quarks, and exotic bound states if they
exist~\cite{Arslandok:2023utm}. Continuous efforts at the theory
frontiers~\cite{An:2021wof,Putschke:2019yrg,Phillips:2020dmw} along with
recent and upcoming plans for state-of-the-art numerical simulations, combined
with sophisticated techniques involving machine learning techniques, may be
able to give us more promising estimates regarding the uncertainties, in
particular the dependence of the uncertainities on the temperature of the
plasma. The main difficulty one meets is that, differently from the
asymptotic freedom regime, in this energy regions the coupling constant of the
strong interactions cannot be used for an application of weak perturbation
techniques. Thus, the most relevant approach seems to be lattice
computations~\cite{Fodor:2001pe,Aoki:2006we,Aoki:2009sc,Bazavov:2009zn,%
deForcrand:2010ys,Borsanyi:2012cr,Bazavov:2017dus,Bonati:2018nut,%
HotQCD:2018pds,Borsanyi:2020fev,Borsanyi:2021sxv,Kahangirwe:2024cny}. However,
this approach is limited by the sign problem that has a high cost in
computational terms~\cite{Splittorff:2007ck,Hsu:2010zza,Aarts:2012yal,%
Nagata:2021ugx}. This problem implies that regions of higher chemical
potential cannot be reached and, presently, it is very difficult to see if the
point where a first order phase transition happens for QCD can ever be
recovered. 

In such a situation, an analytical or semi-analytical approach for QCD at
finite temperature and density is strongly needed. On this track, the recent
work in Ref.~\cite{Lu:2023mkn} (see also refs. therein) should be pointed out
where the authors rely on a minimal model with a numerical solution of the set
of Dyson--Schwinger equations. They are able to obtain good agreement with
lattice data for the critical temperature at finite chemical potential. It
should be pointed out that lattice computations work for a small ratio
$\mu_B/T_c(0)$, evaluating the dependence of $T_c(\mu_B)$ on the chemical
potential through a Taylor series. In the range explored so far, lattice
computations do not show any phase transition point beyond the chiral
crossover.

If a critical end point (CEP) exists in the QCD phase diagram and one is able
to understand the regions of the parameter space involving the chemical
potential allowing for a first-order phase transition, the knowledge of
crossovers will be important milestones for such an experimental endeavour.

Non-perturbative methods are mandatory in this endeavour. Although lattice QCD
investigations have shown the analytic crossover at zero chemical potential     
\cite{Aoki:2006we,Aoki:2009sc,Borsanyi:2010bp,Bazavov:2011nk,%
Bhattacharya:2014ara,Bazavov:2014pvz}, the result suffers from the infamous
fermion sign problem in presence of any (real) chemical potential. This sign
problem means that one has a high cost in computational
terms~\cite{Splittorff:2007ck,Hsu:2010zza,Aarts:2012yal,Nagata:2021ugx}. In
literature, a myriad of methods concerning the extrapolation from zero or
imaginary chemical potential into the real chemical potential region are
exploited, and usually they agree with each other for chemical potentials
$\mu_B/T<2$. However, these too suffer from errors for larger chemical
potential, with the consequence of concrete predictions still eluding.

One of possible ways to capture the effect of large chemical potential is
what is known as the continuum methods, i.e., effective field theory models
and the functional approach. For instance, the Polyakov-loop enhanced
effective models like the Polyakov-loop Nambu-Jona--Lasinio model (PNJL) as
proposed in Refs.~\cite{Fukushima:2003fw,Megias:2004hj,Ratti:2005jh}, or the 
Polyakov-loop quark-meson model (PQM)~\cite{Schaefer:2007pw,Skokov:2010wb,%
Herbst:2010rf}, have been investigated to explore various aspects of the QCD
phase diagram, see e.g.~\cite{Drews:2016wpi,Fukushima:2017csk} for recent
review articles on this topic. Essentially, these methods rely on a simple
chiral effective action which is then added to the Polyakov loop potential.
Such an action, serving as a background that couples to Yang-Mills
interactions, leads in particular to confinement properties of the chiral
dynamics. But as understood in this manner, the effect of the presence of
gluons in the medium cannot be captured, as they are not directly active
degrees of freedom. This becomes possible within functional approaches to QCD.
In particular, Dyson--Schwinger equations (DSE), the functional
renormalisation group, the Hamilton variational approach, and the
Gribov--Zwanziger formalism are some example scenarios where one works with
the quark and gluon degrees of freedom. In this manner one is able to
understand somewhat the phase structure of QCD from the order parameter point
of view, mainly extracted directly from Green's functions of the theory. For
the past several years, these methods involving functionals were able to give
us somewhat a preliminary picture of the QCD phase diagram along with some
characteristic properties of quarks and gluons in the plasma. With such
insights they were influential in understanding the implications of
observables related to QCD thermodynamics, the complex nature of the
transport, and fluctuations of QCD in the framework of DSE and Bethe--Salpeter
equations, see the reviews by Roberts and Schmidt~\cite{Roberts:2000aa} and
Fischer~\cite{Fischer:2018sdj}.

Our aim in this work is to show how a fully analytical approach can be
derived from QCD and implemented to evaluate the critical temperature as a
function of the chemical potential. We use an exact solution to the gluonic
sector of QCD recently obtained in Ref.~\cite{Frasca:2015yva}. This solution
exploits the fact that the vacuum as a Fubini--Lipatov instanton could break
translation invariance~\cite{Fubini:1976jm,Lipatov:1976ny}. However, such a
violation could never be observed, as the Yang--Mills field and its potentials
are never observable and the propagator recovers such a symmetry. We already
applied this idea successfully to the evaluation of the hadron vacuum
polarisation contribution to the $g-2$ of the muon~\cite{Frasca:2021yuu}, and
our result still stands against the most recent evaluations of this quantity
by lattice computations~\cite{Boccaletti:2024guq,FermilabLattice:2024yho} and
the experimental data~\cite{Muong-2:2023cdq,Muong-2:2024hpx}. Based on the same
background, we present a derivation of the critical temperature $T_c(\mu_B)$
by analytical means, and we show good agreement with lattice data, in the
given range of the chemical potential, by using a single parameter that is an
energy scale already characterizing QCD in the regime of asymptotic freedom.

The paper is structured in two main sections: In order to make the paper
self-contained, in Sec.~\ref{sec2} we give a derivation from QCD in the
infrared limit. In Sec.~\ref{sec3}, we introduce temperature and chemical
potential and compare our result for the critical temperature with lattice
data. In Sec.~\ref{sec4}, our conclusions are presented.

\section{QCD in the infrared limit \label{sec2}}

For the sake of completeness, in this section we briefly summarize the
derivation of a non-local Nambu--Jona-Lasinio model. Details are presented
elsewhere~\cite{Frasca:2008zp,Frasca:2011bd,Frasca:2012eq,Frasca:2012iv}.
Compared to the previous analysis, we improve on the form of the propagator
in agreement with the existence of exact solutions of the massless scalar
field~\cite{Frasca:2009bc}.

Starting point for the derivation is the QCD lagrangian
\begin{equation}
{\cal L}_{\rm QCD}=\sum_i\bar\psi^i(i\gamma^\mu D_\mu-m_i)\psi^i
  -\frac14F_{\mu\nu}^aF^{\mu\nu}_a-\frac1{2\xi}(\partial_\mu A^\mu_a)^2.
\end{equation}
Here $A$ represents the gluon field and $\psi$ the QCD quarks, $m$ is the mass
and $i$ are the various flavours of the quark. $\xi$ is the gauge fixing
parameter. The gluon field is coupled minimally via the covariant derivative
$D_\mu=\partial_\mu+ig_sA_\mu^aT_a$, and the field strength tensor
$F_{\mu\nu}=F_{\mu\nu}^aT_a$ is given by $ig_sF_{\mu\nu}=[D_\mu,D_\nu]$,
so that $F_{\mu\nu}^a(x)=\partial_\mu A_\nu^a(x)-\partial_\nu A_\mu^a(x)
-g_sf_{abc}A_\mu^b(x)A_\nu^c(x)$. Written explicitly, the action integral
reads
\begin{eqnarray}
\lefteqn{{\cal S}_{\rm QCD}\ =\ \int d^4x\Bigg[\sum_i\bar\psi^i(x)(i\gamma^\mu
  \partial_\mu-m_i-g_s\gamma^\mu A_\mu^a(x)T_a)\psi^i(x)+\strut}\\&&\strut
  +\frac12A_\mu^a(x)(\eta^{\mu\nu}\dAlem-\partial^\mu\partial^\nu)A_\nu^a(x)
  +\frac1{2\xi}A_\mu^a(x)\partial^\mu\partial^\nu A_\nu^a(x)
  +\strut\nonumber\\&&\strut
  +\frac12g_sf_{abc}\left(\partial_\mu A_\nu^a(x)
  -\partial_\nu A_\mu^a(x)\right)A^\mu_b(x)A^\nu_c(x)
  -\frac14g_s^2f_{abc}f_{cde}A^\mu_a(x)A^\nu_b(x)A_\mu^d(x)A_\nu^e(x).\Bigg]
  \nonumber
\end{eqnarray}
The Euler--Lagrange equations for the gluon fields are given by
\begin{eqnarray}\label{ELA}
0&=&\frac{\delta{\cal S}_{\rm QCD}}{\delta A_\mu^a}
  \ =\ (\dAlem\eta^{\mu\nu}-\partial^\mu\partial^\nu)A_\nu^a
  +\frac1\xi\partial^\mu\partial^\nu A_\nu^a-j^\mu_a+\strut\nonumber\\&&\strut
  +g_sf_{abc}\partial_\nu A^\mu_b A^\nu_c
  +g_sf_{abc}A_\nu^b(\partial^\mu A^\nu_c-\partial^\nu A^\mu_c)
  -g_s^2f_{abc}f_{cde}A_\nu^bA^\mu_dA^\nu_e,\qquad
\end{eqnarray}
where we use $j^\mu_a=g_s\bar\psi\gamma^\mu T_a\psi$ to replace the fermionic
current by a generic one. $T_a$ are the Gell-Mann matrices. Feynman gauge
$\xi=1$ is employed to simplify the equation of motion.

\subsection{Solving the system of Dyson--Schwinger equations}
The Euler--Lagrange equation can be translated to an equation of motion for
the Green functions, constructed from the generating functional $Z[j]$, from
which the system of Dyson--Schwinger equations is derived. We start with
$\langle A_\mu^a(x)\rangle=Z[j]G_{1\mu}^{(j)a}(x)$, and by calculating the
functional derivative with respect to $j^\nu_b(x')$, we continue with
\begin{equation}\label{G1G1}
\langle A_\mu^a(x)A_\nu^b(x')\rangle=Z[j]G_{2\mu\nu}^{(j)ab}(x,x')
  +Z[j]G_{1\mu}^{(j)a}(x)G_{1\nu}^{(j)b}(x').
\end{equation}
Finally, a further functional derivative with respect to $j_\nu^b(x'')$ and
the setting $x=x'=x''$ lead to
\begin{eqnarray}
\lefteqn{\langle A_b^\nu(x)A_\mu^d(x)A_\nu^e(x)\rangle\ =\ Z[j]\Big(
  G_{3\mu\nu b}^{(j)de\nu}(x,x,x)+G_{2\mu b}^{(j)d\nu}(x,x)
  G_{1\nu}^{(j)e}(x)+\strut}\nonumber\\&&\strut
  +G_{1\mu}^{(j)d}(x)G_{2\nu b}^{(j)e\nu}(x,x))
  +G_{1b}^{(j)\nu}(x)G_{2\mu\nu}^{(j)de}(x,x)
  +G_{1b}^{(j)\nu}(x)G_{1\mu}^{(j)d}(x)G_{1\nu}^{(j)e}(x)\Big).\qquad
\end{eqnarray}
Inserting all this into the expectation value of the Euler--Lagrange equation
results in
\begin{eqnarray}\label{EL2}
\lefteqn{\dAlem G_{1\mu}^{(j)a}(x)-j_\mu^a(x)\ =\ g_sf_{abc}\Big\{
  \partial^\nu\Big(G_{2\mu\nu}^{(j)bc}(x,x)
  +G_{1\mu}^{(j)b}(x)G_{1\nu}^{(j)c}(x)\Big)+\strut}\nonumber\\&&\strut
  +\left(\partial_\mu G_{2\nu b}^{(j)c\nu}(x,x)
  -\partial_\nu G_{2\mu b}^{(j)c\nu}(x,x)\right)
  +G_{1b}^{(j)\nu}(x)\left(\partial_\mu G_{1\nu}^{(j)c}(x)
  -\partial_\nu G_{1\mu}^{(j)c}(x)\right)\Big\}
  +\strut\nonumber\\&&\strut\kern-12pt
  -g_s^2f_{abc}f_{cde}\Big\{G_{3\mu\nu b}^{(j)de\nu}(x,x,x)
  +G_{2\mu b}^{(j)d\nu}(x,x)G_{1\nu}^{(j)e}(x)
  +\strut\nonumber\\&&\strut
  +G_{1\mu}^{(j)d}(x)G_{2\nu b}^{(j)e\nu}(x,x)
  +G_{1b}^{(j)\nu}(x)G_{2\mu\nu}^{(j)de}(x,x)
  +G_{1b}^{(j)\nu}(x)G_{1\mu}^{(j)d}(x)G_{1\nu}^{(j)e}(x)\Big\}.\qquad
\end{eqnarray}
Next we calculate the functional derivative with respect to $j^\lambda_h(y)$
to obtain
\begin{eqnarray}
\lefteqn{\dAlem G_{2\mu\lambda}^{(j)ah}(x,y)+i\delta^{ah}\eta_{\mu\lambda}
  \delta^{(4)}(x-y)\ =\ g_sf_{abc}\Big\{
  \left(\partial_\mu G_{3\nu b\lambda}^{(j)c\nu h}(x,x,y)-\partial_\nu
  G_{3\mu b\lambda}^{(j)c\nu h}(x,x,y)\right)+\strut}\nonumber\\&&\strut
  +\partial^\nu\Big(G_{3\mu\nu\lambda}^{(j)bch}(x,x,y)
  +G_{2\mu\lambda}^{(j)bh}(x,y)G_{1\nu}^{(j)c}(x)
  +G_{1\mu}^{(j)b}(x)G_{2\nu\lambda}^{(j)ch}(x,y)\Big)
  +\strut\nonumber\\&&\strut
  +G_{2b\lambda}^{(j)\nu h}(x,y)(\partial_\mu G_{1\nu}^{(j)c}(x)
  -\partial_\nu G_{1\mu}^{(j)c}(x))
  +G_{1b}^{(j)\nu}(x)(\partial_\mu G_{2\nu\lambda}^{(j)ch}(x,y)
  -\partial_\nu G_{2\mu\lambda}^{(j)ch}(x,y))\Big\}
  +\strut\nonumber\\&&\strut\kern-12pt
  -g_s^2f_{abc}f_{cde}\Big\{\Big(G_{4\mu\nu b\lambda}^{(j)de\nu h}(x,x,x,y)
  +G_{3\mu\nu b}^{(j)d\nu h}(x,x,y)G_{1\nu}^{(j)e}(x)
  +G_{2\mu b}^{(j)d\nu}(x,x)G_{2\nu\lambda}^{(j)eh}(x,y)
  +\strut\nonumber\\&&\strut
  +G_{2\mu\lambda}^{(j)dh}(x,y)G_{2\nu b}^{(j)e\nu}(x,x)
  +G_{1\mu}^{(j)d}(x)G_{3\nu b\lambda}^{(j)e\nu h}(x,x,y)\Big)
  +\strut\nonumber\\&&\strut
  +G_{2b\lambda}^{(j)\nu h}(x,y)(G_{2\mu\nu}^{(j)de}(x,x)
  +G_{1\mu}^{(j)d}(x)G_{1\nu}^{(j)e}(x))
  +\strut\nonumber\\&&\strut
  +G_{1b}^{(j)\nu}(x)\Big(G_{3\mu\nu\lambda}^{(j)deh}(x,x,y)
  +G_{2\mu\lambda}^{(j)dh}(x,y)G_{1\nu}^{(j)e}(x)
  +G_{1\mu}^{(j)d}(x)G_{2\nu\lambda}^{(j)eh}(x,y)\Big)\Big\}.
\end{eqnarray}
Using the mapping theorem for Yang--Mills~\cite{Frasca:2007uz,Frasca:2009yp}
with $G^{(2)ab}_{\mu\nu}(x,y)=\delta_{ab} \eta_{\mu\nu}G_2(x-y)$,
$G^{(1)a}_\mu(x)=\eta^a_\mu G_1(x)$ and $j_a^\mu(x)=\eta_a^\mu j(x)$ and
contracting with bases dual with respect to the orthogonality and completeness
relations
\begin{equation}\label{complete}
\eta_\mu^a\eta^\mu_b=-\delta^a_b,\qquad
\eta_\mu^a\eta_\nu^a=-(N_c^2-1)\eta_{\mu\nu}/D,
\end{equation}
leads to the scalar equations
\begin{eqnarray}
\dAlem G_1(x)+N_cg_s^2\Big\{(D-1)G_2(0)G_1(x)+G_1(x)^3\Big\}&=&j(x),
  \label{G1eq}\\
\dAlem G_2(x-y)+(D-1)N_cg_s^2\left(G_2(0)+G_1(x)^2\right)G_2(x-y)
  &=&-i\delta^{(4)}(x-y).\label{G2eq}\qquad
\end{eqnarray}
where $D$ is the space-time dimension. For $\Delta m_G^2:=(D-1)\lambda G_2(0)$
with $\lambda:=N_cg_s^2$, where $N_c$ is the number of colour charges, the
equation of motion for the one-point function $G_1(x)$ reduces to
$(\dAlem+\Delta m_G^2)G_1(x)+\lambda G_1(x)^3=j(x)$. In the following we can
choose $D=4$. The corresponding homogeneous equation is solved by 
\begin{equation}
G_1(x)=\mu\sn(k\cdot x+\theta|\kappa)
  =-i\mu\eta\sum_{m{\rm\,odd}}\sqrt\kappa b_me^{im\eta kx/2},\qquad
b_{2n+1}:=\frac{(-1)^nq^{n+1/2}}{\sqrt\kappa(1-q^{2n+1})},
\end{equation}
where $\kappa:=(\Delta m_G^2-k^2)/k^2$, $q:=\exp(-\pi K(1-\kappa)/K(\kappa))$
and $\eta:=\pi/K(\kappa)$ ($K(\kappa)$ is the complete elliptic integral of
the first kind), with the dispersion relation given by
\begin{equation}
k^2=\Delta m_G^2+\frac12\lambda\mu^2.
\end{equation}
$\mu$ and $\theta$ are integration constants, where $\theta=(1+4N)K(\kappa)$
for an arbitrary integer $N$ is fixed in turn by solving the equation
$(\dAlem+\Delta m_G^2+3\lambda G_1(x)^2)G_2(x-y)=\delta^{(4)}(x-y)$
for the two-point function.
\begin{equation}
K(\kappa)=F(\pi/2|\kappa),\quad\mbox{with\ }
F(\varphi|\kappa)=\int_0^\varphi\frac{d\theta}{\sqrt{1-\kappa\sin^2\theta}}
\end{equation}
is the complete elliptic integral of the first kind, and
$\sn(z|\kappa)=\sin\varphi$ is the corresponding Jacobi elliptic sine,
solution of $z=F(\varphi|\kappa)$. For the two-point Green function one ends
up with the momentum space expression
\begin{equation}
\tilde G_2(p)=\sum_{n=0}^\infty\frac{iB_n}{p^2-m_n^2+i\epsilon},\qquad
  B_n:=\frac{(2n+1)^2\eta^3}{2(1-\kappa)}b_{2n+1},
\end{equation}
with $m_n:=(2n+1)m_0$ and $m_0=\eta\sqrt{k^2}/2$, where in addition we found
that $\sum_{n=0}^\infty B_n=1$. Finally, we arrive at
\begin{equation}
G_2(x-y)=\sum_{n=0}^\infty\int\frac{d^4p}{(2\pi)^4}
  \frac{iB_ne^{-ip(x-y)}}{p^2-m_n^2+i\epsilon}.
\end{equation}
Using this Green function, the general solution of the inhomogeneous
equation~(\ref{G1eq}) reads
\begin{equation}
\phi(x)=G_1(x)+\int G_2(x-y)j(y)dy.
\end{equation}
Note that for the nonlinear differential equation at hand, this holds only
approximately, where the expansion parameter of the approximation is given by
the charge $g_s$ related to the current $j(x)$. Therefore, higher functional
powers of $j$ can be omitted safely.

\subsection{Deriving the non-local Nambu--Jona-Lasinio model}
Returning to the original Lagrange density, the inhomogeneous equation to
be solved is of the shape
\begin{equation}
\dAlem A^\nu_a(x)+\ldots=g_s\sum_i\bar\psi^i(x)\gamma^\nu T_a\psi^i(x).
\end{equation}
The stationary solution is obtained by convolution with the Green function,
\begin{equation}
A^\nu_a(x)=-ig_s\int G_2(x-y)\sum_j\bar\psi^j(y)\gamma^\nu T_a\psi^j(y)d^4y,
\end{equation}
which again can be inserted into the equation of motion for the quark field
to give
\begin{equation}
0=\frac{\partial{\cal L}}{\partial\bar\psi^i}-\partial_\mu
  \frac{\partial{\cal L}}{\partial(\partial_\mu\bar\psi^i)}
  =\frac{\partial{\cal L}}{\partial\bar\psi^i}
  =\left(i\gamma^\mu\partial_\mu-m_i-g_s\gamma^\mu A_\mu^aT_a\right)\psi^i.
\end{equation}
Inserting the stationary solution for the Yang--Mills field one obtains
\begin{equation}
0=(i\gamma^\mu\partial_\mu-m_i)\psi^i(x)-ig_s^2\gamma^\mu T_a\psi^i(x)
  \int\sum_jG_2(x-y)\bar\psi^j(y)\gamma_\mu T_a\psi^j(y)d^4y,
\end{equation}
from which one derives the action integral of the quark flavour dynamics (QFD)
to be
\begin{eqnarray}
{\cal S}_{\rm QFD}&=&\int\sum_i\bar\psi^i(x)(i\gamma^\mu\partial_\mu-m_i)
  \psi^i(x)d^4x+\strut\nonumber\\&&\strut
  -ig_s^2\int\sum_{i,j}\bar\psi^i(x)\gamma^\mu T_a\psi^i(x)G_2(x-y)
  \bar\psi^j(y)\gamma_\mu T_a\psi^j(y)d^4y\,d^4x.
\end{eqnarray}
The currents occurring in the interaction part are coloured. In order to
obtain a Nambu-Jona--Lasinio (NJL) action that describes the interaction
between colourless mesons, we have to perform a Fierz rearrangement. This is a
rearrangement in both the colour and spinor states and
reads~\cite{Vogl:1991qt,Klevansky:1992qe,Buballa:2003qv}
\begin{eqnarray}
\lefteqn{g_s^2\bar\psi^i(x)\gamma^\mu T_a\psi^i(x)\,\bar\psi^j(y)\gamma_\mu
  T^a\psi^j(y)\ =}\\
  &=&\frac{N_c^2-1}{2N_c^2}g_s^2\Big(\bar\psi^i(x)\psi^j(y)\,\bar\psi^j(y)
  \psi^i(x)+\bar\psi^i(x)i\gamma_5\psi^j(y)\,\bar\psi^j(y)i\gamma_5
  \psi^i(x)+\strut\nonumber\\&&\strut\qquad
  -\frac12\bar\psi^i(x)\gamma_\mu\psi^j(y)\,\bar\psi^j(y)\gamma^\mu
  \psi^i(x)-\frac12\bar\psi^i(x)\gamma_5\gamma_\mu\psi^j(y)\,
  \bar\psi^j(y)\gamma_5\gamma^\mu\psi^i(x)\Big)+\strut\nonumber\\&&\strut
  -\frac{g_s^2}N_c\Big(\bar\psi^i(x)T_a\psi^j(y)\,\bar\psi^j(y)T_a\psi^i(x)
  -\bar\psi^i(x)\gamma_5T_a\psi^j(y)\,\bar\psi^j(y)\gamma_5T_a\psi^i(x)
  +\strut\nonumber\\&&\strut\qquad
  -\frac12\bar\psi^i(x)\gamma_\mu T_a\psi^j(y)\,\bar\psi^j(y)\gamma^\mu
  T_a\psi^i(x)-\frac12\bar\psi^i(x)\gamma_5\gamma_\mu T_a\psi^j(y)\,
  \bar\psi^j(y)\gamma_5\gamma^\mu T_a\psi^i(x)\Big).\nonumber
\end{eqnarray}
The result consists of a singlet and an octet contribution. Because of the
minus sign, the latter is repulsive and will not contribute to a meson bound
state. In addition, the limit $N_c\to\infty$ will make it vanish. Therefore,
we keep only this first part that again contains scalar, pseudoscalar, vector,
and axial vector currents. As the (two) flavour states are now mixed, we can
divide $\bar\psi^i(x)\psi^j(y)$ up into the isoscalar current
$\bar\psi^i(x)\delta_{ij}\psi^j(y)$ and the isovector current
$\bar\psi^i(x)\vec\sigma_{ij}\psi^j(y)$ where
$\vec\sigma=(\sigma_1,\sigma_2,\sigma_3)\in SU(2)_f$. Finally, we take out the
scalar--isoscalar and the pseudoscalar--isovector currents as those
representing physical mesons, and combine them into a four-vector current
$\bar\psi(x)\Gamma^\alpha\psi(y)$ (employing $\Gamma^0=\bbbone$ and
$\Gamma^i=\gamma_5\sigma_i=-\Gamma_i$) with
$\bar\psi(y)\Gamma^\alpha\psi(x)=(\bar\psi(x)\Gamma^\alpha\psi(y))^*$ to
obtain
\begin{eqnarray}
\lefteqn{{\cal S}_{\rm NJL}\ =\ \int\bar\psi(x)(i\gamma^\mu\partial_\mu
  -\hat m)\psi(x)d^4x+\strut}\nonumber\\&&\strut
  -i\frac{N_c^2-1}{2N_c^2}g_s^2\int\bar\psi(x)\Gamma_\alpha\psi(y)G_2(x-y)
  \bar\psi(y)\Gamma^\alpha\psi(x)d^4y\,d^4x,
\end{eqnarray}
where $\hat m=\mbox{diag}\{m_i\}$ is the constituent quark mass matrix in
flavour space. Using the relative coordinates $w=(x+y)/2$ and $z=x-y$
($x=w+z/2$, $y=w-z/2$), one replaces the two-point Green function by
$G_2(z)=-iG{\cal C}(z)/2$ with $G=2/((1-\kappa)k^2)$. It is easy to show that
$\tilde{\cal C}(0)=\int{\cal C}(z)d^4z=1$, a detail that is postponed to
Appendix~A.

\subsection{Performing the bosonisation procedure}
For the interaction part of the NJL action, one obtains~\cite{Hell:2008cc}
\begin{equation}
{\cal S}_{\rm int}=-\frac{GG_S^2}2\int\bar\psi(w+\frac z2)\Gamma_\alpha
  \psi(w-\frac z2){\cal C}(z)\bar\psi(w-\frac z2)\Gamma^\alpha
  \psi(w+\frac z2)d^4w\,d^4z
\end{equation}
with $G_S^2:=(N_c^2-1)g_s^2/(2N_c^2)$. Including
${\cal S}_0=\int\bar\psi(x)(i\gamma^\mu\partial_\mu-\hat m)\psi(x)d^4x$, the
functional integral reads ${\cal Z}_{\rm NJL}=\int{\cal D}\bar\psi{\cal D}\psi
\exp(i({\cal S}_0+{\cal S}_{\rm int}))$. In order to perform the
bosonisation, one adds a mesonic field $(\phi^\alpha)=(\sigma,\vec\pi)$ by
multiplying the factor
\begin{equation}
{\cal N}^{-1}=\int{\cal D}\phi^*{\cal D}\phi\exp\left(\frac{i}{2G}
  \int{\cal C}(z)\phi_\alpha^*(w)\phi^\alpha(w)d^4w\,d^4z\right),
\end{equation}
to the functional integral, leading to an action ${\cal S}_{\rm int+}$. Note
that the integration over $z$ is actually trivial. Still, one needs this
integration to perform the functional shift
\begin{equation}
\phi_\alpha(w)\to\phi_\alpha(w)-GG_S\bar\psi(w-\frac z2)\Gamma_\alpha
  \psi(w+\frac z2).
\end{equation}
In performing this shift, the quartic interaction cancels and one obtains
the action integral
\begin{eqnarray}
\lefteqn{{\cal S}_{\rm int+}[\psi,\bar\psi,\phi,\phi^*]\ =\ \int\Bigg[
  \frac1{2G}{\cal C}(z)\phi_\alpha^*(w)\phi^\alpha(w)
  +\strut}\nonumber\\&&\strut\kern-1pt
  -\frac{G_S{\cal C}(z)}2\Big(\phi_\alpha(w)\bar\psi(w+\frac z2)\Gamma^\alpha
  \psi(w-\frac z2)+\phi_\alpha^*(w)\psi(w-\frac z2)\Gamma^\alpha
  \psi(w+\frac z2)\Big)\Bigg]d^4w\,d^4z\ =\nonumber\\
  &=&\frac1{2G}\int\phi_\alpha^*(w)\phi^\alpha(w)d^4w
  -G_S\int\bar\psi(x){\cal C}(x-y)\real\phi_\alpha\pfrac{x+y}2
  \Gamma^\alpha\psi(y)d^4x\,d^4y,\qquad
\end{eqnarray}
where one has returned to the previous coordinates, though for the second term
in reverse order (note that ${\cal C}(x-y)={\cal C}(y-x)$). The main
bosonisation process consists in integrating out the fermionic fields. This is
done by performing a Fourier transform to
\begin{eqnarray}
\lefteqn{{\cal S}_{\rm int+}[\psi,\bar\psi,\phi,\phi^*]
  \ =\ \frac1{2G}\int\frac{d^4q}{(2\pi)^4}\tilde\phi_\alpha^*(q)
  \tilde\phi^\alpha(q)+\strut}\nonumber\\&&\strut
  -\frac{G_S}2\int\frac{d^4p'}{(2\pi)^4}\frac{d^4p}{(2\pi)^4}
  \tilde{\bar\psi}(p')\tilde{\cal C}\pfrac{p'+p}2\left(\tilde\phi_\alpha(p'-p)
  +\tilde\phi_\alpha(p-p')\right)\Gamma^\alpha\tilde\psi(p),
\end{eqnarray}
combining ${\cal S}_0$ with the second part of ${\cal S}_{\rm int+}$, and
performing the calculation
\begin{eqnarray}
\lefteqn{\int{\cal D}\bar\psi{\cal D}\psi\exp\Bigg[i\int\frac{d^4p'}{(2\pi)^4}
  \frac{d^4p}{(2\pi)^4}\tilde{\bar\psi}(p')\times\strut}\\&&\strut
  \kern-24pt\times\left((2\pi)^4\delta^{(4)}(p'-p)(\gamma^\mu p_\mu-\hat m)
  -\frac{G_S}2\tilde{\cal C}\pfrac{p'+p}2\left(\tilde\phi_\alpha(p'-p)
  +\tilde\phi_\alpha(p-p')\right)\Gamma^\alpha\right)\tilde\psi(p)\Bigg]
  \ =\kern-9pt\nonumber\\
  &=&\det\left((2\pi)^4\delta^{(4)}(p'-p)(\gamma^\mu p_\mu-\hat m)
  -\frac{G_S}2\tilde{\cal C}\pfrac{p'+p}2\left(\tilde\phi_\alpha(p'-p)
  +\tilde\phi_\alpha(p-p')\right)\Gamma^\alpha\right).\nonumber
\end{eqnarray}
The matrix of which the determinant is taken is understood as being expressed
not only in terms of Dirac and (trivial) colour and flavour matrices but also
between states $|p\rangle$ and $|p'\rangle$. The bosonisation procedure is
completed by the mean field approximation $\phi(z)=(\bar\sigma,\vec 0)$,
resulting in
$\phi_\alpha(p'-p)=(2\pi)^4\delta^{(4)}(p'-p)\bar\sigma\delta_{\alpha0}$. This
renders the trace over states a single four-dimensional integration. For the
determinant one obtains
\begin{equation}
\det\left((2\pi)^4\delta^{(4)}(p'-p)(\gamma^\mu p_\mu-\hat M_f(p))\right)
  =\int\frac{d^4p}{(2\pi)^4}\prod_f(p^2-M_f^2(p))^{2N_c},
\end{equation}
where $\hat M_f(p)=\hat m_f+G_S\tilde{\cal C}(p)\bar\sigma\bbbone$ is the gap
mass matrix in flavour space with diagonal elements
$M_f(p)=m_f+G_S\tilde{\cal C}(p)\bar\sigma$. Therefore, the bosonised action
integral is given by
\begin{equation}
{\cal S}_{\rm bos}=\frac{\bar\sigma^2}{2G}
  -2iN_c\int\frac{d^4p}{(2\pi)^4}\sum_f\ln(p^2-M_f^2(p)),
\end{equation}
In calculating the variation with respect to $\bar\sigma$ one obtains
\begin{equation}
\bar\sigma=-4iN_cGG_S\int\frac{d^4p}{(2\pi)^4}\sum_f
  \frac{\tilde{\cal C}(p)M_f(p)}{p^2-M_f^2(p)},
\end{equation}
and inserting this into the formula for $M_f(p)$, one ends up with the mass
gap equation
\begin{equation}\label{massgapp}
M_q(p)=m_q-4iN_cGG_S^2\tilde{\cal C}(p)\int\frac{d^4q}{(2\pi)^4}\sum_f
  \frac{\tilde{\cal C}(q)M_f(q)}{q^2-M_f^2(q)}.
\end{equation}
For massless quarks, the sum over the flavours is performed trivially and
results in
\begin{equation}
M(p)=-4iN_fN_cGG_S^2\tilde{\cal C}(p)\int\frac{d^4q}{(2\pi)^4}
  \frac{\tilde{\cal C}(q)M(q)}{q^2-M^2(q)}.
\end{equation}

\section{QCD at finite temperature  \label{sec3}}

\subsection{Introducing the temperature}
In the following, we deal with massless quarks. At finite temperature, the
integration over the zeroth momentum component is replaced by a Matsubara sum,
resulting in
\begin{equation}
M(i\omega_k,\vec p\,)=4N_fN_cGG_S^2\tilde{\cal C}(i\omega_k,\vec p\,)
  \beta^{-1}\sum_{l=-\infty}^\infty\int\frac{d^3q}{(2\pi)^3}
  \frac{\tilde{\cal C}(i\omega_l,\vec q\,)M(i\omega_l,\vec q\,)}{\omega_l^2
  +\vec q\,^2+M^2(i\omega_l,\vec q\,)}.
\end{equation}
Using $M(i\omega_k,\vec p\,)=\tilde{\cal C}(i\omega_k,\vec p\,)G_S\bar\sigma$,
the mass gap equation can be rewritten into a mass gap equation for
$\bar\sigma$,
\begin{equation}\label{massgaps}
1=4N_fN_cGG_S^2\beta^{-1}\sum_{k=-\infty}^{+\infty}\int\frac{d^3p}{(2\pi)^3}
  \frac{\tilde{\cal C}^2(i\omega_k,\vec p\,)}{\omega_k^2+\vec p\,^2
  +\tilde{\cal C}^2(i\omega_k,\vec p\,)G_S^2\bar\sigma^2}=:f_M(\beta^{-1}).
\end{equation}
Using $N_f=3$, $N_c=3$, $\omega_k=(2k+1)\omega_0$, $\omega_0=\pi/\beta$,
\begin{equation}
G=\frac2{(1-\kappa)k^2}=\frac{2K(\kappa)^2}{(1-\kappa)\pi^2m_0^2},\qquad
G_S^2=\frac{N_c^2-1}{2N_c^2}g_s^2=\frac494\pi\alpha_s
\end{equation}
(note that for $\kappa=-1$, $G$ is equal to the inverse string tension), and
\begin{equation}\label{tildeC}
\tilde{\cal C}(i\omega_k,\vec p\,)=\sum_{n=0}^\infty
  \frac{m_0^2C_{2n+1}(\kappa)}{\omega_k^2+\vec p\,^2+(2n+1)^2m_0^2},\qquad
C_\nu(\kappa)=\frac{2\pi\nu^2}{K(\kappa)}\
  \frac{(-1)^nq^{\nu/2}}{\sqrt\kappa(1-q^\nu)}=C_{-\nu}(\kappa)
\end{equation}
with nome $q=\exp(-\pi K(1-\kappa)/K(\kappa))$ (again for $\kappa=-1$, one has
$q=\exp(-(1-i)\pi)$ and, therefore,
$C_\nu(-1)=(2\pi\nu^2/K(-1))e^{\pi\nu/2}/(1+e^{\pi\nu})=C_{-\nu}(-1)$).

Eq.~(\ref{massgaps}) can be rendered as a dimensionless equation by using
$\hat m_0=\beta m_0/2$, $\hat\omega_0=\beta\omega_0/2=\pi/2$ and
$\hat\rho=\beta\rho/2$. As such, the equation is no longer dependent on
the temperature and on the ground state Matsubara frequency. The mass gap
equation reads
\begin{equation}\label{massgaphat}
1=\frac{N_fN_c\hat GG_S^2}{\pi^2}\sum_{k=-\infty}^{+\infty}\int_0^\infty
  \frac{\tilde C^2(i\hat\omega_k,\hat\rho)\hat\rho^2d\hat\rho}
  {\hat\omega_k^2+\hat\rho^2+\tilde C^2(i\hat\omega_k,\hat\rho)\hat\sigma^2}.
\end{equation}
with $\hat\sigma:=\beta G_S\bar\sigma/2$,
\begin{equation}
\hat\omega_k=(2k+1)\hat\omega_0=\frac12(2k+1)\pi,\qquad
\hat G=\frac{2K(\kappa)^2}{(1-\kappa)\pi^2\hat m_0^2}=\frac{4G}{\beta^2}.
\end{equation}

\subsection{Including the quark chemical potential}
Now we can include the quark chemical potential $\mu_f$. For this we start with
the grand canonical potential
\begin{equation}
\Omega=-\frac1{2\beta}\sum_{k=-\infty}^{+\infty}\int\frac{d^3p}{(2\pi)^3}\Tr\ln
  \left[\beta\tilde S^{-1}(i\omega_k;\vec p\,)\right]+\frac{\bar\sigma^2}{2G}
\end{equation}
with an additional factor $1/2$ because of the doubling of the degrees of
freedom in the inverse Nambu--Gor'kov propagator
\begin{equation}
\tilde S^{-1}(i\omega_k;\vec p\,)=\begin{pmatrix}\left((i\omega_k\bbbone_c-A_4)
  \bbbone_f+\hat\mu_f\bbbone_c\right)\gamma^0-\vec\gamma\cdot\vec p\,
  \bbbone_{c,f}-\hat M(i\omega_k,\vec p\,)\bbbone_d\quad 0\quad\\
  \quad 0\quad\left((i\omega_k^*\bbbone_c-A_4)\bbbone_f-\hat\mu_f\bbbone_c\right)
  \gamma^0-\vec\gamma\cdot\vec p\,\bbbone_{c,f}-\hat M^*(i\omega_k,\vec p\,)
  \bbbone_d\\\end{pmatrix}.
\end{equation}
Using $\Tr\ln=\ln\det$, the next step is to calculate the (huge) determinant
of this inverse propagator multiplied by $\beta$. Using the fact that the
determinant of a block diagonal matrix is the product of the determinants of
the blocks, one obtains 
\begin{eqnarray}
\lefteqn{\det(\beta\tilde S^{-1})\ =\ \det\left\{\beta\left[\left(i(\omega_k
  \bbbone_c-A_4)\bbbone_f+\hat\mu_f\bbbone_c\right)\gamma^0
  -\vec\gamma\cdot\vec p\,\bbbone_{c,f}-\hat M(i\omega_k,\vec p\,)\bbbone_d
  \right]\right\}\cdot\strut}\nonumber\\&&\strut\cdot
  \det\left\{\beta\left[\left(i(\omega_k\bbbone_c-A_4)\bbbone_f-\hat\mu_f
  \bbbone_c\right)\gamma^0-\vec\gamma\cdot\vec p\,\bbbone_{c,f}
  -\hat M^*(i\omega_k,\vec p\,)\bbbone_d\right]\right\}\ =\nonumber\\
  &=&\prod_f\det\left\{\beta\left[\left(i(\omega_k\bbbone_c-A_4)+\mu_f\bbbone_c
  \right)\gamma^0-\vec\gamma\cdot\vec p\,\bbbone_c-\hat M(i\omega_k,\vec p\,)
  \bbbone_d\right]\right\}\cdot\strut\nonumber\\[-7pt]&&\strut\qquad\cdot
  \det\left\{\beta\left[\left(i(\omega_k\bbbone_c-A_4)-\mu_f\bbbone_c
  \right)\gamma^0-\vec\gamma\cdot\vec p\,\bbbone_c-\hat M^*(i\omega_k,\vec p\,)
  \bbbone_d\right]\right\}\ =\nonumber\\
  &=&\prod_f\det\left\{\beta\left[i\omega_{k,f}^-\gamma^0-\vec\gamma\cdot\vec
  p\,-M(i\omega_k,\vec p\,)\bbbone_d\right]\right\}
  \det\left\{\beta\left[i\omega_{k,f}^{-*}\gamma^0-\vec\gamma\cdot\vec
  p\,-M^*(i\omega_k,\vec p\,)\bbbone_d\right]\right\}
  \cdot\kern-6pt\strut\nonumber\\[-12pt]&&\strut\cdot
  \det\left\{\beta\left[i\omega_{k,f}^+\gamma^0-\vec\gamma\cdot\vec
  p\,-M(i\omega_k,\vec p\,)\bbbone_d\right]\right\}
  \det\left\{\beta\left[i\omega_{k,f}^{+*}\gamma^0-\vec\gamma\cdot\vec
  p\,-M^*(i\omega_k,\vec p\,)\bbbone_d\right]\right\}
  \cdot\strut\nonumber\\&&\strut\kern-12pt\cdot
  \det\left\{\beta\left[i\omega_{k,f}^0\gamma^0-\vec\gamma\cdot\vec
  p\,-M(i\omega_k,\vec p\,)\bbbone_d\right]\right\}
  \det\left\{\beta\left[i\omega_{k,f}^{0*}\gamma^0-\vec\gamma\cdot\vec
  p\,-M^*(i\omega_k,\vec p\,)\bbbone_d\right]\right\}\ =\nonumber\\
  &=&\prod_f\left[\beta^2(\omega_{k,f}^-)^2+\beta^2\vec p\,^2
  +\beta^2M^2(i\omega_k,\vec p\,)\right]^2
  \left[\beta^2(\omega_{k,f}^{-*})^2+\beta^2\vec p\,^2
  +\beta^2M^{*2}(i\omega_k,\vec p\,)\right]^2
  \cdot\strut\nonumber\\[-12pt]&&\strut\cdot
  \left[\beta^2(\omega_{k,f}^+)^2+\beta^2\vec p\,^2
  +\beta^2M^2(i\omega_k,\vec p\,)\right]^2
  \left[\beta^2(\omega_{k,f}^{+*})^2+\beta^2\vec p\,^2
  +\beta^2M^{*2}(i\omega_k,\vec p\,)\right]^2
  \cdot\strut\nonumber\\&&\strut\kern-8pt\cdot
  \left[\beta^2(\omega_{k,f}^0)^2+\beta^2\vec p\,^2
  +\beta^2M^2(i\omega_k,\vec p\,)\right]^2
  \left[\beta^2(\omega_{k,f}^{0*})^2+\beta^2\vec p\,^2
  +\beta^2M^{*2}(i\omega_k,\vec p\,)\right]^2,\qquad
\end{eqnarray}
where the three Matsubara frequencies $\omega_k^-$, $\omega_k^+$ and
$\omega_k^0$ are the three diagonal components of the colour space matrix
$\hat\omega_k:=\omega_k\bbbone_c-A_4$ with only diagonal colour matrices taken
into account, resulting in $\omega_k^\pm=\omega_n\pm A_4^3/2-A_4^8/(2\sqrt3)$,
$\omega_k^0=\omega_k+A_4^8/\sqrt3$, and the quark chemical potentials are
absorbed into the Matsubara frequencies by defining
$\omega_{k,f}^\lambda:=\omega_k^\lambda-i\mu_f$, $\lambda=\pm,0$. For the last
step we have used that $\det[\beta(\gamma^0 p^0-\vec\gamma\cdot\vec p\,
-m\bbbone_d)]\ =$
\begin{equation}
\ =\ \begin{vmatrix}\beta(p^0-m)&0&-\beta p^3&-\beta p^1+i\beta p^2\\
  0&\beta(p^0-m)&-\beta p^1-i\beta p^2&\beta p^3\\
  \beta p^3&\beta p^1-i\beta p^2&\beta(-p^0-m)&0\\
  \beta p^1+i\beta p^2&-\beta p^3&0&\beta(-p^0-m)\\\end{vmatrix}
  \ =\ (-\beta^2p^2+\beta^2m^2)^2.\qquad
\end{equation}
Calculating the logarithm of $\det(\beta\tilde S^{-1})$, one has
\begin{eqnarray}
\lefteqn{\Omega\ =\ \frac{\bar\sigma^2}{2G}-\frac1\beta\sum_f
\sum_{k=-\infty}^{+\infty}  \int\frac{d^3p}{(2\pi)^3}
  \times\strut}\nonumber\\&&\strut\times\Big[
  \ln\left(\beta^2(\omega_{k,f}^-)^2+\beta^2\vec p\,^2
  +\beta^2M^2(i\omega_k,\vec p\,)\right)
  +\ln\left(\beta^2(\omega_{k,f}^{-*})^2+\beta^2\vec p\,^2
  +\beta^2M^{*2}(i\omega_k,\vec p\,)\right)+\strut\nonumber\\&&\strut
  +\ln\left(\beta^2(\omega_{k,f}^+)^2+\beta^2\vec p\,^2
  +\beta^2M^2(i\omega_k,\vec p\,)\right)
  +\ln\left(\beta^2(\omega_{k,f}^{+*})^2+\beta^2\vec p\,^2
  +\beta^2M^{*2}(i\omega_k,\vec p\,)\right)+\strut\nonumber\\&&\strut\kern-8pt
  +\ln\left(\beta^2(\omega_{k,f}^0)^2+\beta^2\vec p\,^2
  +\beta^2M^2(i\omega_k,\vec p\,)\right)
  +\ln\left(\beta^2(\omega_{k,f}^{0*})^2+\beta^2\vec p\,^2
  +\beta^2M^{*2}(i\omega_k,\vec p\,)\right)\Big].\qquad
\end{eqnarray}
Taking into account that
$M(i\omega_k,\vec p\,)=\tilde{\cal C}(i\omega_k,\vec p\,)G_S\bar\sigma$, one
can perform the variation with respect to $\bar\sigma$ as usual, and in
minimising $\Omega$, one obtains the mass gap equation
\begin{equation}
1=\frac{\hat GG_S^2}{2\pi^2}\sum_f\sum_{k=-\infty}^{+\infty}
  \sum_{\lambda=\pm,0}\int_0^\infty
  \left[\frac{\tilde C^2(i\hat\omega_k,\hat\rho)\hat\rho^2
  d\hat\rho}{(\hat\omega_{k,f}^\lambda)^2+\hat\rho^2
  +\hat{\cal C}^2(i\hat\omega_k,\hat\rho)\hat\sigma^2}
  +\frac{\tilde C^2(i\hat\omega_k,\hat\rho)\hat\rho^2
  d\hat\rho}{(\hat\omega_{k,f}^{\lambda*})^2+\hat\rho^2
  +\hat{\cal C}^2(i\hat\omega_k,\hat\rho)\hat\sigma^2}\right],
\end{equation}
where the hatted propagator $\hat{\cal C}(i\hat\omega_k,\hat\rho)$ is given
by Eq.~(\ref{tildeC}) with all dimensional quantities replaced by
dimensionless (hatted) quantities. Note that the right hand side of the mass
gap equation that we call the {\em mass gap function\/} is real. With $A_4=0$
(no outer field), $\hat\omega_{k,f}^\lambda=\hat\omega_k-i\hat\mu_f$ does no
longer depend on the colour label $\lambda$. Therefore, the sum over $\lambda$
will result in a factor $N_c$. One obtains
\begin{equation}\label{massgap}
1=\frac{N_c\hat GG_S^2}{\pi^2}\sum_f\sum_{k=-\infty}^{+\infty}\int_0^\infty
  \frac{\left(\hat\omega_k^2-\hat\mu_f^2+\hat\rho^2+\hat{\cal C}^2
    (i\hat\omega_k,\hat\rho)\hat\sigma^2\right)\hat{\cal C}^2
    (i\hat\omega_k,\hat\rho)}{\left(\hat\omega_k^2-\hat\mu_f^2+\hat\rho^2
    +\hat{\cal C}^2(i\hat\omega_k,\hat\rho)\hat\sigma^2\right)^2
    +4\hat\omega_k^2\hat\mu_f^2}\hat\rho^2d\hat\rho.
\end{equation}
Taking all quark chemical potentials to be equal, the sum over the flavours
$f$ can be replaced by $N_f$. For our standard choice $N_f=3$ and $N_c=3$ we
obtain the solution of the mass gap equation in dependence on $\hat\sigma$ for
different values of the reduced chemical potential $\hat\mu_f=\beta\mu_f/2$.
In principle, using the solution of the mass gap equation one can describe
the critical temperature in dependence on the chemical potential. In order to
see this, one has to get back to dimensional quantities. Depending on $m_0$ as
a constant, the critical temperature can be calculated as
$T_c=m_0/2\hat m_0$. On the other hand, the dimensional quark chemical
potential is given by $\mu_f=2T_c\hat\mu_f$. Therefore, the abscissa have to
be scaled with $2T_c$, i.e., the result of the calculation. The best way to
show this dependence is via a parametric plot that also suggests that the
dependence $T_c(\mu_f)$ is not a function. 

\subsection{Adjusting to lattice data}
\begin{figure}
\begin{center}
\epsfig{figure=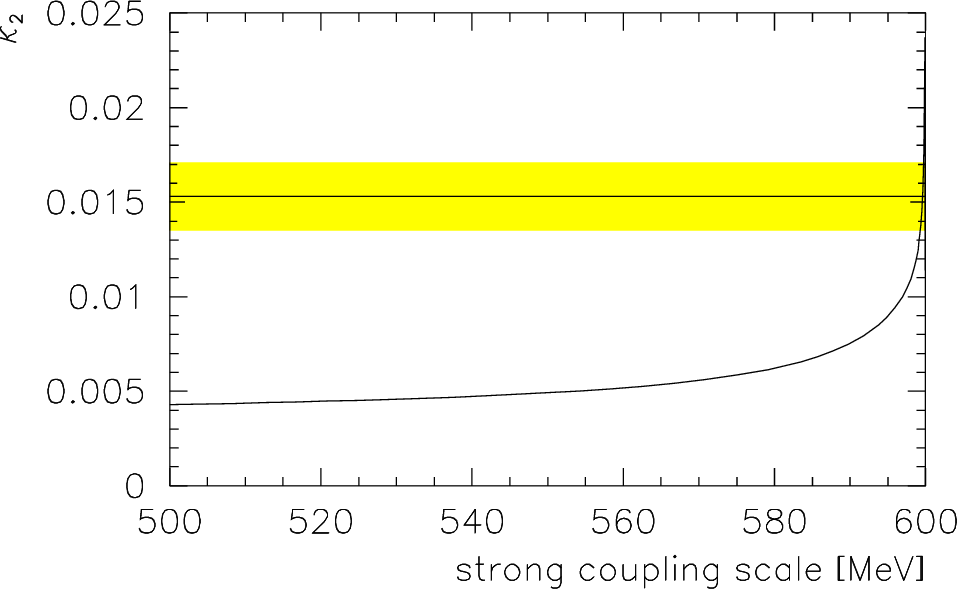, scale=0.8}
\epsfig{figure=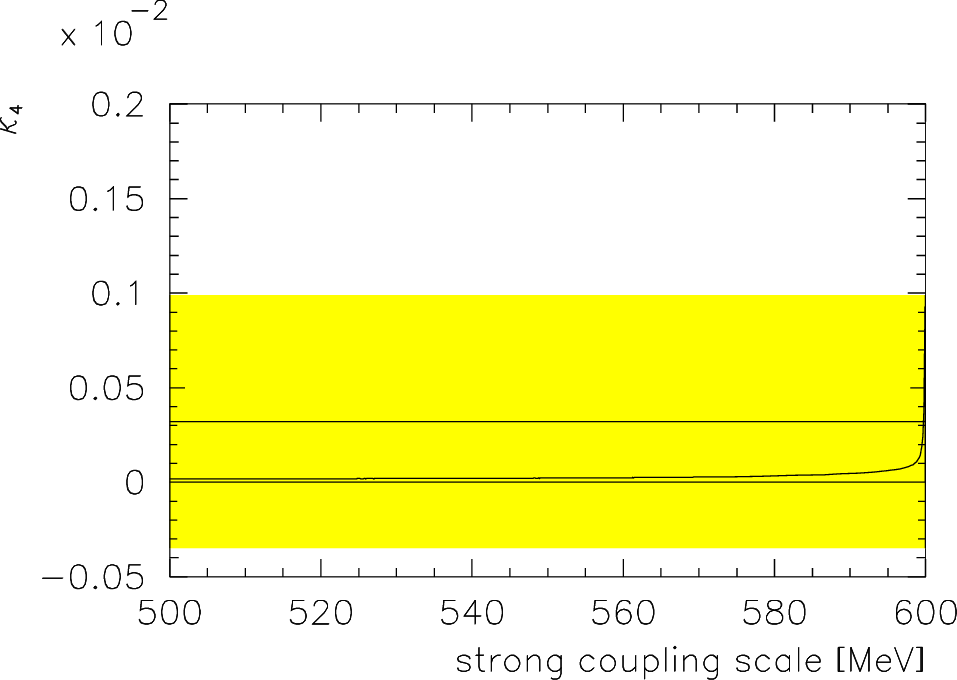, scale=0.8}
\caption{\label{kappa2al}Values for $\kappa_2$ (upper panel) and $\kappa_4$
(lower panel) in dependence on the strong coupling scale, as compared to the
values from lattice calculations (yellow band with central
line, values taken from Ref.~\cite{Borsanyi:2020fev})}
\end{center}\end{figure}

Before we can calculate the parametric plot, we have to adjust the only free
parameter of the model, namely $m_0$. For this we use lattice results. In
Ref.~\cite{Borsanyi:2020fev}, a detailed analysis of the critical temperature
in dependence on the baryon chemical potential has been performed close to the
point $\mu_B=0$. The normalised critical temperature is seen to be an even
function of the chemical potential and can be expanded in a Taylor
series,\footnote{The Taylor series expansion shown here is actually given in
Ref.~\cite{Lu:2023mkn}. Note that in Ref.~\cite{Borsanyi:2020fev} the
chemical potential was normalised to the critical temperature at the baryon
chemical potential, not at the chemical potential at zero. The difference is
marginal, though.}
\begin{equation}
\frac{T_c(\mu_B)}{T_c(0)}=1-\frac{\kappa_2\mu_B^2}{T_c(0)^2}
  -\frac{\kappa_4\mu_B^4}{T_c(0)^4}+\ldots
\end{equation}
Taking the right hand side of the mass gap equation~(\ref{massgap}) as a
function of $\hat\mu_f^2$ and $\hat m_0^2$, the mass gap equation
$F(\hat\mu_f^2,\hat m_0^2)=1$ can be understood as an implicit equation that
related the two variables in a functional way, at least close to $\mu_B=0$.
Our first approach to the Taylor coefficients $\kappa_2$ and $\kappa_4$ is a
numerical one. Approximating the solution of the mass gap equation
$\hat m_0^2(\hat\mu_f^2)$ in dependence on $\hat\mu_f^2$ by a polynomial up to the
power of two in $\hat\mu_f^2$, one can extract the value $\hat m_0^2$ and the
derivatives $(\hat m_0^2)'$ and $(\hat m_0^2)''$ at $\hat\mu_f^2=0$. On the
other hand, the functional dependence one aims to determine the Taylor
coefficients for is the one of
$f(\hat\mu_f^2):=T_c(\hat\mu_f^2)/T_c(0)=\hat m_0(0)/\hat m_0(\hat\mu_f^2)$ on
$g(\hat\mu_f^2):=(\mu_B/T_c(0))^2=(3\mu_f/T_c(0))^2=36\hat\mu_f^2\hat m_0(0)^2/\hat
m_0(\hat\mu_f^2)^2$. Calculating iteratively
\begin{equation}
\frac{df}{dg}=\frac{f'}{g'},\qquad
\frac{d^nf}{dg^n}=\frac{d}{dg}\pfrac{d^{n-1}f}{dg^{n-1}}
  =\frac1{g'}\pfrac{d^{n-1}f}{dg^{n-1}},
\end{equation}
where the prime indicates derivative with respect to $\hat\mu_f^2$, one can
reach up to arbitrary high Taylor coefficients for $f(g)$. In this way, we
obtain
\begin{equation}
\kappa_2=\frac{(\hat m_0^2)'}{72\hat m_0^2},\qquad\kappa_4=\frac{2(\hat
m_0^2)''\hat m_0^2+((\hat m_0^2)')^2}{2!(72\hat m_0^2)^2},\qquad\ldots
\end{equation}
Having obtained $\hat m_0^2=0.374757^2=0.140443$, $(\hat m_0^2)'=0.04333$ and
$(\hat m_0^2)''=0.0055$, one ends up with $\kappa_2=0.00429$ and
$\kappa_4=0.000017$. These values are roughly one third of the values given in
Ref.~\cite{Borsanyi:2020fev}. However, the choice of the scale of the strong
coupling, originally chosen at $500\MeV$, can now be used to adjust our
prediction to the lattice data. A first sketch unveils that this adjustment is
indeed possible. In Fig.~\ref{kappa2al} we have shown the numerical values for
$\kappa_2$ (upper panel) and $\kappa_4$ (lower panel) in dependence on the
scale of the strong coupling in the interval between $500$ and $600\MeV$.
Shown are also the lattice results including the error bar. It is obvious that
close to the right boundary of $600\MeV$, there is a chance to match our
result to the lattice data.

\begin{figure}[H]\begin{center}
\epsfig{figure=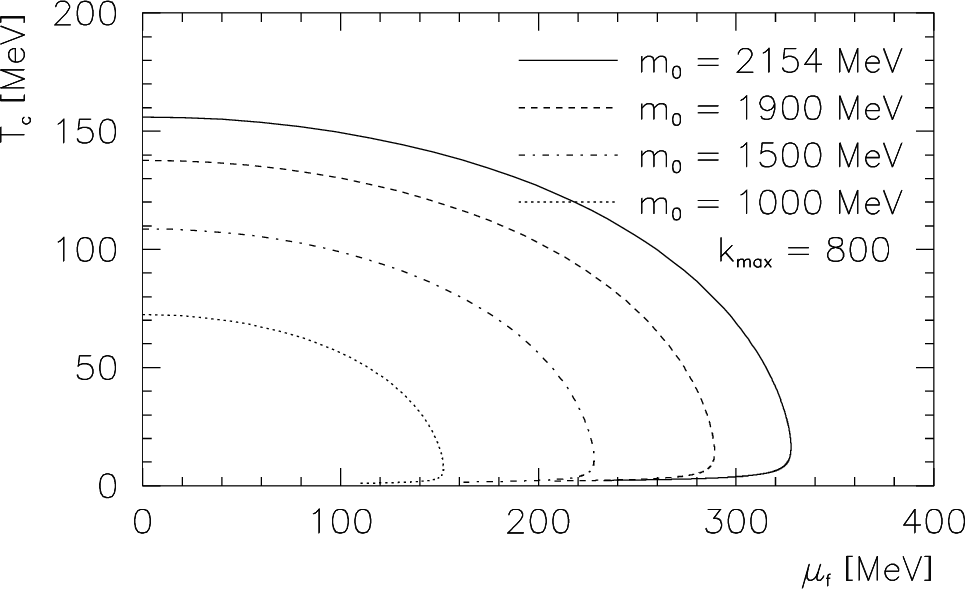, scale=0.8}
\caption{\label{critem800}Dependence of the critical temperature $T_c$ on the
quark chemical potential $\mu_f$. As stated in the text, we evaluate the
running coupling through its value in the asymptotic regime, extending the
validity at the energy scale fitted to the lattice.}

\end{center}\end{figure}

The matching can be also done semi-analytically. For this we return to the
mass gap function $F(x,y)$ as an implicit function. Taking partial derivatives
with respect to the two arguments $x=\hat\mu_f^2$ and $y=\hat m_0^2$, we obtain
\begin{eqnarray}
0&=&\frac{\partial F}{\partial x}+\frac{\partial F}{\partial y}\frac{dy}{dx},
  \nonumber\\
0&=&\frac{\partial^2F}{\partial x^2}
  +2\frac{\partial^2F}{\partial x\partial y}\frac{dy}{dx}
  +\frac{\partial^2F}{\partial y^2}\pfrac{dy}{dx}^2
  +\frac{\partial F}{\partial y}\frac{d^2y}{dx^2},
\end{eqnarray}
from which we derive
\begin{eqnarray}\label{dm0dmu}
\frac{dy}{dx}&=&-\frac{\partial F}{\partial x}
  \pfrac{\partial F}{\partial y}^{-1},\nonumber\\
\frac{d^2y}{dx^2}&=&-\left(\frac{\partial^2F}{\partial x^2}
  +2\frac{\partial^2F}{\partial x\partial y}\frac{dy}{dx}
  +\frac{\partial^2F}{\partial y^2}\pfrac{dy}{dx}^2\right)
  \pfrac{\partial F}{\partial y}^{-1}\ =\nonumber\\
  &=&-\left(\frac{\partial^2F}{\partial x^2}\pfrac{\partial F}{\partial y}^2
  -2\frac{\partial^2F}{\partial x\partial y}\frac{\partial F}{\partial x}
  \frac{\partial F}{\partial y}+\frac{\partial^2F}{\partial y^2}
  \pfrac{\partial F}{\partial x}^2\right)\pfrac{\partial F}{\partial y}^{-3}.
  \qquad
\end{eqnarray}
Via a procedure consisting of different steps that is described for the mass
gap function in detail in Appendix~B but works in the same way also for the
derivatives, one can perform the sum over the Matsubara frequencies explicitly
and replace the integration over $\hat\rho$ by a sum over residues. Together
with the two-fold sum over the mass states $m_n=(2n+1)m_0$ in
Eq.~(\ref{tildeC}) of the two propagator factors
$\hat{\cal C}(\hat\omega_k,\hat\rho)$, one is left with three-fold summations
for
\begin{eqnarray}
F\Big|_{\hat\mu_f^2=0}&=&-\frac{F_0}8\sum_{\nu,\nu_1,\nu_2{\rm\ odd}}
  \frac{2\pi C_{\nu_1}(\kappa)C_{\nu_2}(\kappa)}{\nu_1^2-\nu_2^2}
  \times\strut\nonumber\\&&\strut\times\Bigg(\frac{\sqrt{4\nu_1^2\hat
  m_0^2+\nu^2\pi^2}-|\nu|\pi}{4\nu_1^2\hat m_0^2}-\frac{\sqrt{4\nu_2^2\hat
  m_0^2+\nu^2\pi^2}-|\nu|\pi}{4\nu_2^2\hat m_0^2}\Bigg),\nonumber\\
\frac{\partial F}{\partial\hat\mu_f^2}\Bigg|_{\hat\mu_f^2=0}
  &=&\frac{F_0}8\sum_{\nu,\nu_1,\nu_2{\rm\ odd}}
  \frac{8\pi C_{\nu_1}(\kappa)C_{\nu_2}(\kappa)}{\nu_1^2-\nu_2^2}
  \times\strut\nonumber\\&&\strut\times\Bigg(
  \frac{\sqrt{(4\nu_1^2\hat m_0^2+\nu^2\pi^2}-|\nu|\pi)^3}{(4\nu_1^2
    \hat m_0^2)^3}
  -\frac{\sqrt{(4\nu_2^2\hat m_0^2+\nu^2\pi^2}-|\nu|\pi)^3}{(4\nu_2^2
  \hat m_0^2)^3}\Bigg),\nonumber\\
\frac{\partial F}{\partial\hat m_0^2}\Bigg|_{\hat\mu_f^2=0}
  &=&\frac{F_0}8\sum_{\nu,\nu_1\nu_2{\rm\ odd}}
  \frac{\pi C_{\nu_1}(\kappa)C_{\nu_2}(\kappa)}{\hat m_0^2(\nu_1^2-\nu_2^2)}
  \times\strut\nonumber\\&&\strut\times\Bigg(
  \frac{(\sqrt{4\nu_1^2\hat m_0^2+\nu^2\pi^2}-|\nu|\pi)^2}{4\nu_1^2\hat m_0^2
  \sqrt{4\nu_1^2\hat m_0^2+\nu^2\pi^2}}
  -\frac{(\sqrt{4\nu_2^2\hat m_0^2+\nu^2\pi^2}-|\nu|\pi)^2}{4\nu_2^2\hat m_0^2
  \sqrt{4\nu_2^2\hat m_0^2+\nu^2\pi^2}}\Bigg),\nonumber\\
\frac{\partial^2F}{\partial(\hat\rho^2)^2}\Bigg|_{\hat\mu_f^2=0}
  &=&-\frac{F_0}8\sum_{\nu,\nu_1\nu_2{\rm\ odd}}
  \frac{64\pi C_{\nu_1}(\kappa)C_{\nu_2}(\kappa)}{\nu_1^2-\nu_2^2}
  \times\strut\nonumber\\&&\strut\times\Bigg(
  \frac{(\sqrt{4\nu_1^2\hat m_0^2+\nu^2\pi^2}-|\nu|\pi)^5}{(4\nu_1^2
  \hat m_0^2)^5}
  -\frac{(\sqrt{4\nu_2^2\hat m_0^2+\nu^2\pi^2}-|\nu|\pi)^5}{(4\nu_2^2
  \hat m_0^2)^5}\Bigg),\nonumber\\
\frac{\partial^2F}{\partial\hat\rho^2\partial\hat m_0^2}
  \Bigg|_{\hat\mu_f^2=0}&=&-\frac{F_0}8\sum_{\nu,\nu_1\nu_2{\rm\ odd}}
  \frac{12\pi C_{\nu_1}(\kappa)C_{\nu_2}(\kappa)}{\hat m_0^2
  (\nu_1^2-\nu_2^2)}\times\strut\nonumber\\&&\strut\times\Bigg(
  \frac{(\sqrt{4\nu_1^2\hat m_0^2+\nu^2\pi^2}-|\nu|\pi)^4}{(4\nu_1^2
  \hat m_0^2)^3\sqrt{4\nu_1^2\hat m_0^2+\nu^2\pi^2}}
  -\frac{(\sqrt{4\nu_2^2\hat m_0^2+\nu^2\pi^2}-|\nu|\pi)^4}{(4\nu_2^2
  \hat m_0^2)^3\sqrt{4\nu_2^2\hat m_0^2+\nu^2\pi^2}}\Bigg),\nonumber\\
\frac{\partial^2F}{\partial(\hat m_0^2)^2}
  \Bigg|_{\hat\mu_f^2=0}&=&-\frac{F_0}8\sum_{\nu,\nu_1\nu_2{\rm\ odd}}
  \frac{\pi C_{\nu_1}(\kappa)C_{\nu_2}(\kappa)}{2\hat m_0^4(\nu_1^2-\nu_2^2)}
  \times\strut\nonumber\\&&\strut\times\Bigg(
  \frac{(\sqrt{4\nu_1^2\hat m_0^2+\nu^2\pi^2}-|\nu|\pi)^3
  (3\sqrt{4\nu_1^2\hat m_0^2+\nu^2\pi^2}+|\nu|\pi)}{4\nu_1^2
  \hat m_0^2(4\nu_1^2\hat m_0^2+\nu^2\pi^2)^{3/2}}
  +\strut\kern-36pt\nonumber\\&&\strut
  -\frac{(\sqrt{4\nu_2^2\hat m_0^2+\nu^2\pi^2}-|\nu|\pi)^3
  (3\sqrt{4\nu_2^2\hat m_0^2+\nu^2\pi^2}+|\nu|\pi)}{4\nu_2^2
  \hat m_0^2(4\nu_2^2\hat m_0^2+\nu^2\pi^2)^{3/2}}\Bigg).\qquad
\end{eqnarray}
In order to begin with the numerical analysis and to match the lattice results,
we start with the value of the general factor $F_0$ that contains the strong
coupling, for $\kappa=-1$ given by
\begin{equation}
F_0=\frac{N_fN_c\hat GG_S^2}{\pi^2}
  =3\times 3\times\frac{K(-1)^2}{\pi^2}\times\frac{16\pi\alpha_s}{9\pi^2}
  =\frac{16}{\pi^3}K(-1)^2\alpha_s=0.886941\alpha_s.
\end{equation}
The Taylor coefficient $\hat m_0(\hat\mu_f^2=0)$ is solution of
$F(0,\hat m_0^2)=1$. The dependence of $F(0,\hat m_0)$ shows a saturation for
high values of $\hat m_0$ at approximately $0.28F_0=0.25\alpha_s$. Therefore,
in order that the mass gap equation is satisfied, $\alpha_s$ has to be larger
than $4.0$. In our case at the perturbative scale of $500\MeV$, we have
$\hat m_0=0.374757$. Inserting this value into the derivatives and using these
to calculate the slope and the curvature of $\hat m_0^2(\hat\mu_f^2)$
according to Eq.~(\ref{dm0dmu}), one obtains
\begin{equation}
(\hat m_0^2)'=\frac{d(\hat m_0^2)}{d(\hat\mu_f^2)}\Bigg|_{\hat\mu_f^2=0}
  =0.0433332,\qquad
(\hat m_0^2)''=\frac{d^2(\hat m_0^2)}{d(\hat\mu_f^2)^2}\Bigg|_{\hat\mu_f^2=0}
  =0.0110326,
\end{equation}
that are in nice agreement with the previous numerical values, or improve
their precision (note that the calculation of the second derivative very much
depends on the mesh length of the approximation). The resulting coefficients
are given by $\kappa_2=0.00429$ and $\kappa_4=0.000024$. In order to adjust to
the lattice values $\kappa_2=0.0153(18)$ and $\kappa_4=0.00032(67)$ found in
Ref.~\cite{Borsanyi:2020fev}, we have to solve
$(\hat m_0^2)'=72\hat m_0^2\kappa_2$, or
\begin{equation}
\frac{\partial F}{\partial\hat\mu_f^2}+72\kappa_2\hat m_0^2
  \frac{\partial F}{\partial\hat m_0^2}=0,
\end{equation}
which is an implicit equation for $\hat m_0$. For $\kappa_2=0.0153$, the
matching procedure gives $\hat m_0=7.7503$. For this value, one has
$F(0,\hat m_0^2)=0.278128F_0$ and $\kappa_4=0.000314$. The value of the
coupling corresponding to this is $\alpha_s=4.05379$. The scale corresponding
to the strong coupling is $m_0=599.56\MeV$, where we used four-loop running
with matching at flavour thresholds and the value
$\alpha_s(m_Z)=0.1175\textstyle\substack{+0.0025\\-0.0028}$
for $m_Z=91.1876\pm 0.0021\GeV$. From the theoretical point of view, we notice
that we are working in a regime where the perturbative expression for the
running coupling in the regime of asymptotic freedom reaches its limit of
applicability. We extend this limit by taking the result it gives at its face
value, recognizing that we are in a deeply non-perturbative regime. Indeed,
with the current knowledge of the running coupling in a strong
coupled regime, this should be considered just another fitting parameter and
our choice arise from pure consistency reasons. Thus, taking the strong
coupling $\alpha_s$ at the scale of $599.56\MeV$, the parametric plot is shown
in Figure~\ref{critem800} for the values $m_0=1000\MeV$, $1500\MeV$,
$2000\MeV$ and $2157\MeV$, the latter close to the expectation.

\section{Discussion and Conclusions}\label{sec4}

Using a closed-form solution for the correlation functions of the Yang--Mills
theory, we show how to derive a non-local Nambu--Jona-Lasinio model directly
from QCD, describing the behaviour of the theory away from the asymptotic
freedom regime. This model has also been proven to be
confining~\cite{Frasca:2022lwp} and to be able to give results for the hadron
vacuum polarisation correction to the muon's $g-2$ factor in agreement with
the experiment~\cite{Frasca:2021yuu}. In this work, we extend the
applicability of this model in regimes of finite temperature and chemical
potential where lattice data are available. We were able to show that:
\begin{itemize}
\item The theory displays a phase transition in agreement with the lattice
  data results, yielding the crossover point of the chiral phase transition.
\item We evaluate the coefficients $\kappa_2$ and $\kappa_4$ of the Taylor
  expansion of $T_c(\mu_B)$ around $\mu_B=0$ and find agreement with lattice
  data in a large range of values of our QCD scale $m_0$ common to both
  coefficients. This is shown in Fig.~\ref{kappa2al} where the yellow bands
  show the agreement zone with lattice data as a function of $m_0$. 
\item We derive the critical temperature $T_c$ as a function of the chemical
  potential $\mu_f$ in Eq.~(\ref{massgap}). In Fig.~\ref{critem800}, we show
  such a dependence at varying a single parameter representing the proper
  scale of the model (the QCD scale $m_0$). This parameter arises from the
  integration of the Yang-Mills theory, determines the spectrum of the theory,
  and yields the mass gap. We are able to obtain an excellent agreement with
  lattice data related to the physical scale we are ranging on.
\end{itemize}
We aim to obtain the QCD equation of state in our future works. Due to the
possible dynamics coming from the quark--gluon plasma and the electroweak
transitions, they impact the propagation of primordial gravitational waves
(PGWs) in the early universe~\cite{Schwarz:1997gv,Maggiore:1999vm,%
Mazumdar:2018dfl}, with prospects for measurements in current and upcoming
gravitational wave (GW) experiments like DECIGO~\cite{Seto:2001qf,Sato:2017dkf},
LISA~\cite{Audley:2017drz}, SKA~\cite{Janssen:2014dka}, and
EPTA~\cite{Lentati:2015qwp} across various frequency ranges with quantified
estimates shown in Ref.~\cite{Hajkarim:2019csy}. The results obtained will
have large impact for the understanding of such GW measurements. Besides this,
we envisage also to have applications of our results in precise estimates of
relic density of dark matter in the early universe and its experimental direct
detection~\cite{Drees:2015exa,Steigman:2012nb},  and cosmological phase
transition with observable effects~\cite{Capozziello:2018qjs,Khodadi:2021ees}.
For detailed analyses on these topics see Refs.~\cite{Drees:2015exa,
Schwarz:1997gv,Castorina:2018whj,Anand:2017kar,Schwarz:2003du,Stuke:2011wz,%
Saikawa:2018rcs,Hindmarsh:2005ix,Borsanyi:2016ksw,Capozziello:2018qjs,%
Li:2018oqf}. In particular, the effect of considering finite non-zero chemical
potential, as it is studied in this paper, may modify the strength of the
cosmological transition with inevitable consequences for the early universe if
it undergoes a first or second order phase
transition~\cite{Witten:1984rs,Asakawa:1989bq}. This may also lead to a
difference in the equation of state when compared to the case of zero chemical
potentials~\cite{Schwarz:2009ii,Stuke:2011wz}. Such cosmological analyses are
beyond the scope of the current paper and we leave these to future work.

\begin{appendix}

\section{On the normalised two-point Green function ${\cal C}(x)$}
\setcounter{equation}{0}\def\theequation{A\arabic{equation}}
The normalised two-point Green function is given by ${\cal C}(z)=2iG_2(x)/G$,
where $G$ is obtained in turn by calculating
$\tilde G_2(0)=\int G_2(z)d^4z=-iG/2$. This is easy to do. Indeed, on the one
hand one has
\begin{equation}
\sn(\zeta|\kappa)=\frac{2\pi}{K(\kappa)\sqrt\kappa}\sum_{n=0}^\infty
  \frac{q^{n+1/2}}{1-q^{2n+1}}\sin\left((2n+1)\frac{\pi\zeta}{2K(\kappa)}
  \right),
\end{equation}
and for $\zeta=K(\kappa)$
\begin{equation}
1=\sn(K(\kappa)|\kappa)=\frac{2\pi}{K(\kappa)}\sum_{n=0}^\infty
  \frac{q^{n+1/2}}{\sqrt\kappa(1-q^{2n+1})}\sin\left((2n+1)\frac\pi2\right)
  =\frac{2\pi}{K(\kappa)}\sum_{n=0}^\infty
  \frac{(-1)^nq^{n+1/2}}{\sqrt\kappa(1-q^{2n+1})}.
\end{equation}
On the other hand,
\begin{equation}
\tilde G_2(0)=\sum_{n=0}^\infty\frac{iB_n}{-m_n^2}
  =\frac{-2\pi i}{(1-\kappa)K(\kappa)k^2}\sum_{n=0}^\infty
  \frac{(-1)^nq^{n+1/2}}{\sqrt\kappa(1-q^{2n+1})}
  =\frac{-i}{(1-\kappa)k^2}=-\frac{iG}2.
\end{equation}
Therefore, one has $G=2/((1-\kappa)k^2)$.

\section{The mass gap function and its derivatives}
\setcounter{equation}{0}\def\theequation{B\arabic{equation}}
Starting point is the mass gap function
\begin{equation}
F(\hat\mu_f^2,\hat m_0^2)=\frac{F_0}{\hat m_0^2}\sum_{k=-\infty}^{+\infty}
  \int_0^\infty\frac{\left(\hat\omega_k^2+\hat\rho^2-\hat\mu_f^2\right)
  \tilde{\cal C}^2(i\hat\omega_k,\hat\rho)}{\left(\hat\omega_k^2+\hat\rho^2
  -\hat\mu_f^2\right)^2+4\hat\omega_k^2\hat\mu_f^2}\hat\rho^2d\hat\rho
  =:F(\hat\mu_f^2,\hat m_0^2)
\end{equation}
with $F_0:=N_cN_f\hat GG_S^2/\pi^2$. The first step is the introduction of
Feynman parameters, according to
\begin{equation}
\frac1{A_1^{\alpha_1}\cdots A_m^{\alpha_m}}
  =\frac{\Gamma(\alpha_1+\ldots+\alpha_m)}{\Gamma(\alpha_1)\cdots
  \Gamma(\alpha_m)}\int_0^1\frac{x_1^{\alpha_1-1}\cdots x_m^{\alpha_m-1}
  \delta(x_1+\ldots+x_m-1)}{(x_1A_1+\ldots+x_mA_m)^{\alpha_1+\ldots+\alpha_m}}
  dx_1\cdots dx_m,
\end{equation}
while the second step is the explicit summation over the Matsubara frequencies,
\begin{eqnarray}
\lefteqn{F(\hat\mu_f^2,\hat m_0^2)\ =\ F_0\sum_{k=-\infty}^{+\infty}
  \sum_{n_1,n_2=0}^\infty\int_0^\infty\frac{\hat m_0^2C_{2n_1+1}(\kappa)
  C_{2n_2+1}(\kappa)\hat\rho^2d\hat\rho}{(\hat\omega_k^2+\hat\rho^2
  +(2n_1+1)^2\hat m_0^2)(\hat\omega_k^2+\hat\rho^2+(2n_2+1)^2\hat m_0^2)
  (\hat\omega_k^2+\hat\rho^2)}\ =}\nonumber\\
  &=&F_0\sum_{k=-\infty}^{+\infty}\sum_{n_1,n_2=0}^\infty\int_0^\infty
  \int_0^1dx_1\int_0^{1-x_1}dx_2\frac{\Gamma(3)C_{2n_1+1}(\kappa)
  C_{2n_2+1}(\kappa)\hat m_0^2\hat\rho^2d\hat\rho}{\left(\hat\omega_k^2
  +\hat\rho^2+x_1(2n_1+1)^2\hat m_0^2+x_2(2n_2+1)^2\hat m_0^2\right)^3}
  \ =\kern-18pt\nonumber\\
  &=&F_0\sum_{n_1,n_2=0}^\infty C_{2n_1+1}(\kappa)C_{2n_2+1}(\kappa)
  \int_0^1dx_1\int_0^{1-x_1}dx_2\times\strut\nonumber\\&&\strut\times
  \int_0^\infty\frac{\hat m_0^2\hat\rho^2d\hat\rho}{4\hat a^5}
  \Big[2\hat a^2\tanh^3\hat a+3\hat a\tanh^2\hat a
  +(3-2\hat a^2)\tanh\hat a-3\hat a\Big]
\end{eqnarray}
with $\hat a^2=\hat\rho^2+\hat m^2$ and
$\hat m^2=x_1(2n_1+1)^2\hat m_0^2+x_2(2n_2+1)^2\hat m_0^2$. Applying Cauchy's
residue theorem, the integrand provides poles up to degree $5$ at
$\hat\rho=\pm i\hat m$ from the general factor and poles up to degree $3$ at
$\hat\rho=\pm i\sqrt{\hat m^2+(2n+1)^2\pi^2/4}$ from the hyperbolic tangent
functions. While the former residues related to the first poles vanish, the
residues of the latter are given by
\begin{equation}
\frac{\mp 2i}{(4\hat m^2+(2n+1)^2\pi^2)^{3/2}}.
\end{equation}
One obtains
\begin{eqnarray}
\lefteqn{F(\hat\mu_f^2,\hat m_0^2)\ =\ F_0\sum_{n,n_1,n_2=0}^\infty\int_0^1dx_1
  \int_0^{1-x_1}dx_2\times\strut}\nonumber\\&&\strut
  \frac{2\pi C_{2n_1+1}(\kappa)C_{2n_2+1}(\kappa)\hat m_0^2}{\left(4x_1
  (2n_1+1)^2\hat m_0^2+4x_2(2n_2+1)^2\hat m_0^2+(2n+1)^2\pi^2\right)^{3/2}}.
\end{eqnarray}
Finally, the integrations over the Feynman parameters $x_1$ and $x_2$ can be
performed to give
\begin{eqnarray}
\lefteqn{F(\hat\mu_f^2,\hat m_0^2)=\frac{F_0}8\sum_{\nu,\nu_1,\nu_2{\rm\ odd}}
  \frac{2\pi C_{\nu_1}(\kappa)C_{\nu_2}(\kappa)}{\nu_1^2-\nu_2^2}
  \times\strut}\nonumber\\&&\strut\times\left(
  \frac{\sqrt{4\nu_2^2\hat m_0^2+\nu^2\pi^2}-|\nu|\pi}{4\nu_2^2\hat m_0^2}
  -\frac{\sqrt{4\nu_1^2\hat m_0^2+\nu^2\pi^2}-|\nu|\pi}{4\nu_1^2\hat m_0^2}
  \right).
\end{eqnarray}
In this expression we have used $\nu=2n+1$, $\nu_1=2n_1+1$ and $\nu=2n+1$ and
symmetrised the summation, giving rise to the factor $1/8$. The partial
derivatives of this mass gap function are handled in the same way. Note that
in calculating $\kappa_2$ and $\kappa_4$, the general factor $F_0$ cancels out.
Only the original mass gap equation depends in the value of $F_0$.

\end{appendix}

\end{document}